# A FRAMEWORK FOR MANAGING COST EFFECTIVE AND EASY ELECTRONIC PAYMENT SYSTEM IN THE DEVELOPING COUNTRIES

Asif Ahmed Anik[1] and Al-Mukaddim Khan Pathan[2]

[1]Department of Information and Communication Engineering, The University of Tokyo.
E-mail: anik@logos.ic.i.u-tokyo.ac.jp

[2]Department of Computer Science and Information Technology (CIT), Islamic University of Technology (IUT), Board Bazar, Gazipur-1704, Bangladesh.
E-mail: al_muka@yahoo.com

**Abstract**

With the rapid growth of Information and Communication Technology, Electronic commerce is now acting as a new means of carrying out business transactions through electronic means such as Internet environment. To avoid the complexities associated with the digital cash and electronic cash, consumers and vendors are looking for credit card payments on the Internet as one possible time-tested alternative. This gave rise of the on-line payment processing using a third-party verification; which is not suitable for the developing countries in most of the cases because of the excessive costs associated with it for maintenance and establishment of an online third-party processor. As a remedy of this problem, in this paper, we have proposed a framework for easy security incorporation in credit card based electronic payment system without the use of an on-line third- party processor; which tends to be low cost and effective for the developing countries.

## 1. INTRODUCTION

In the past, business was pretty simple. Suppose, a person having a pig and another person having some wool showed up at the market and haggled, and then one went home to knit a cardigan while the other may prepare to arrange some barbecue.

Now-a-days this type of show up at the market is not possible, and moreover, people don't even need physical goods or currency to conduct business. Electronic commerce is the most recent step in the evolution of business transactions. It replaces (or augments) the swapping of money or goods with the exchange of information from computer to computer [1].

Though the electronic commerce sounds like a great opportunity with low overhead, few employees and no physical location, several questions are needed to be addressed to resolve the issue of payment. Electronic commerce and electronic business greatly need new payment systems that will support their further development [2]. On-line electronic payment systems are widely used in e-commerce and include wholesale payments, wire transfers, recurring bill payments, the automated clearinghouse, electronic draft capture, and electronic check presentment [3]. On-line electronic commerce payments include token based payment systems and credit card-based payment systems.

An on-line electronic payment system using encrypted credit card suffers from several problems including security, risks and financial unattractiveness for microtransactions. While, a credit card-based payment system using third-party verification asks for the use of a third-party (i.e., a company that collects and approves payments from one client to another) on the Internet to verify the electronic transactions incurring extra costs. Using such a third-party in electronic payment systems, may not be suitable for the developing countries due to the complexity and excessive cost associated with it; which may make it unpopular. Since such a hurdle is not insurmountable; in order to resolve this issue in the developing countries, in this paper, a suitable electronic payment system is proposed, which does not require an on-line third-party connection. More importantly it is low cost with necessary security incorporation thereby suitable for the developing countries.

The rest of the paper is organized as follows: Section 2 gives an overview of the existing credit card-based electronic payment systems, Section 3 identifies the issues to be considered to design an electronic payment system, Section 4 deals with the idea behind the proposed electronic payment system, Section 5 elaborates the architecture, Section 6 enlighten the advantages of the proposed framework, Section 7 reflects the implementation issues and Section 8 concludes the paper.

## 2. EXISTING CREDIT CARD-BASED ELECTRONIC PAYMENT SYSTEMS

Though the use of credit cards during electronic transaction added a new flavor in electronic commerce, there is nothing new in the basic process. The consumers, who want to buy a product or service, simply send their credit card details to the involving service provider and the credit card organization handles this payment like any other electronic transaction [3]. Existing credit card-based electronic payment systems include: (a) the use of encrypted credit cards (e.g., World Wide Web form-based encryption) and (b) third-party authorization (e.g., First Virtual).

### 2.1. Encryption and credit cards [3]

In this scheme, in order to make a truly secure and nonrefutable transaction using an encrypted credit card, each consumer and each vendor generates a public and a secret key. The public key is sent to the credit card company and put on its public key server. The secret key is re-encrypted with a password and the unencrypted version is erased. To buy something from vendor X, the consumer sends vendor X the message, "It is now time $T$, I am paying $Y$ dollars to $X$ for item $Z$," then the consumer uses his or her password to sign the message with the public key. The vendor will then sign the message with its own secret key and send it to the credit card company, which will bill the consumer for $Y$ dollars and give the same amount (less a fee) to $X$.

### 2.2 Third-party processors and credit cards [3] [4]

A third party credit card processor is a company that accepts credit card orders on behalf of other online businesses. In third-party processing, consumers register with a third party on the Internet to verify the electronic microtransactions. Here, the two key servers are merchant server and payment server. Using a client browser, a user makes a purchase from a merchant server by clicking on a payment URL (hyperlinks), which is attached to the product on a WWW page. The payment URLs send the encoded information including the details of purchase (e.g., price of item, target URL and duration) to the payment server. If the information entered by the customer is valid and funds are available, the payment server processes the payment transaction and redirects the user's browser to the purchased item with an access URL, which encodes the details of the payment transaction (the amount, what was purchased and duration).

The access URL acts as a digital invoice, stamped "paid" by the payment server; which provides evidence to the merchant that the user has paid for the information and provides a receipt that grants the user access. The merchant runs an HTTP server that is modified to process access URLs. The server checks the validity of the URL and grants access if the expiration time has not passed. Once the customer is authenticated, the payment is automatically processed.

**2.3. Drawbacks of the existing credit card-based electronic payment systems**

Though the existing credit card-based electronic payment systems are simple in comparison to the digital cash and electronic checks, they suffer from a variety of drawbacks. The encrypted credit card-based electronic payment system suffers from the following disadvantages:

- The credit card companies need to maintain a public server with all the public keys assuming that the credit card company will keep the vendor honest.
- Encrypted credit card transactions may not be micro enough for purchasing information and numerous half-dollar and one dollar transactions may not be financially attractive, compared to the average credit card transaction of about $60.
- If the encrypted credit card electronic payment system is extended to all of the small-dollar services, available over the internet (e.g., 20-cent file transfers and $1 video game rentals), the overall processing load on key system components will likely become unmanageable or commercially nonviable unless a significant amount of automation takes place.
- The companies maintaining credit-card based payment systems have to be big enough so that the costs for management and maintenance of the system do not entail considerable profit lose. Technological and financial strength of the company need to be pretty solid.

The electronic payment system using a third-party processor dominates over the encrypted credit card-based system for microtransactions. But it also has disadvantages, stated as follows:

- Requiring an on-line third-party connection for each transaction to different banks could lead to processing bottlenecks that can undermine the goal of reliable use.
- To use this system both the customers and merchants must be registered with the On-line third-party processor (OTTP). In the case of First Virtual, this registration costs $2 for buyers and $10 for sellers. Sellers also pay a fee of 29 cents for each transaction plus 2 percent. Sellers also pay a $1 processing fee when aggregated payments are made to their account [3].
- The complexity of credit card processing using an on-line third-party, takes place in the verification phase; a potential bottleneck. Verification using an OTTP is time consuming and may require many sequence-specific operations. This may lead the system at stake.
- In the third-world developing countries, electronic commerce is not considered as a lucrative one and the general thought of the people is to consider it as an expensive one and rather unprofitable. Under these circumstances, using an OTTP for credit card verification is not wise in the third-world developing countries; since the establishment and the maintenance of the OTTP requires a large amount of funding. Moreover, payment of the fee to the third-party both from the buyers and the sellers may make the people and the merchants reluctant of using the OTTP for credit card verification scheme.
- Third-party agents are required to be trustful. Otherwise, using a third-party for credit card verification may seem a bit risky since the transactions are not anonymous and credit card companies do in fact compile valuable data about spending habits.
- Since an OTTP is a centralized entity, it is a candidate for single point of failure. If the third-party processor collapses, all ongoing transactions may be hampered, causing inconvenience to the users.

## 3. CONSIDERATIONS TO DESIGN AN ON-LINE ELECTRONIC PAYMENT SYSTEM

Despite the cost and efficiency gains, many hurdles remain in building an on-line electronic payment system (EPS). These include several factors, addressed as below:

- Privacy should be assured as the users expect to trust a secure system. Electronic communication through the use of electronic payment system should be as safe as a private medium like a telephone free of wiretaps and hackers.
- Although no systems are yet fool-proof, electronic payment systems designers should concentrate closely on security.
- As the users value convenience more than anything, the payment interface should be user friendly having intuitive outlook and must be as easy to use as a telephone.
- Designing an EPS should handle the challenge to integrate the databases used by each of the users, while keeping the data up-to-date and error free.
- A "network broker"- someone to broker goods and services, settle conflicts and facilitate financial transactions electronically- must be in place [3].
- An EPS should resolve the issue of how to price payment system services. The necessity of using subsidies to price all service affordably should be recognized and the potential waste of resources for using subsidies to encourage users to shift from one form of payment to another should be considered.
- A common standard should be imposed and followed; since without standards the wielding of different payment users into different networks and different systems is impossible [3].
- An EPS should be cheaper both for the buyers and sellers. Minimum cost should be associated with establishing an on-line electronic payment system.

Keeping these factors in mind, we have thought for a new framework suitable for use in the developing countries where costing is a great issue. In most of the developing countries, the e-commerce and electronic business are in nascent stage mainly due to the costing. Moreover, the users of the developing countries are not that much technologically sound to understand, handle and believe the credit card transaction often. Hence, we are proposing the new framework which is not only easy to incorporate but also cost effective.

## 4. THE PROPOSED ON-LINE CARD-BASED ELECTRONIC PAYMENT SYSTEM

With a view to overcome the above stated problems associated with encrypted credit card and on-line third-party processor, in this paper, we have proposed the framework for a new electronic payment system for easy security incorporation; which does not use an on-line third party processor and which is also suitable for business implementation in the third world developing countries.

## 4.1. The idea behind

In the third world developing countries, there has been little use of e-commerce applications. Even different e-commerce applications like the electronic payment system is not that much popular in those countries. The companies and the e-commerce service providers always look for their profit. But since the cost for establishment and maintenance of the on-line third-party processor is considerably high, they are not interested to use it and as a matter of fact, the credit card-based electronic payment system using on-line third-party processor is not at all used by them. So, in our system we have proposed the framework for a credit card-based electronic payment system that makes no use of the on-line third-party processor. Rather it limits the transaction within only two parties.

In the proposed system, there are two parties associated with a transaction. One party includes- the credit card providers (e.g., VISA, MasterCard), charge card providers (e.g., American Express), debit card providers (e.g., Bank accounts), digital card providers and private label card providers (J. C. Penny) and most importantly even the Internet Service Providers who sell prepaid cards for dialup internet access. Other party is the service provider for making on-line transactions. The consumers can get cards from the card providers who can eventually make these cards available in different stores and shops.

All the companies taking part in the card provider consortium can prepare and sell cards. But they should maintain a fixed format. Each card provider has a unique id. A card contains information about- the card provider id, secret number/text, balance etc. Each of the companies will also maintain a database on the sold cards from where the credits will be made. Moreover, each company will maintain a server which will listen on a defined port for the 'Credit Requests'. Generally the card providers will prepare the card and will make them widely available in different shops. The cards will be ranging from very low amounts to high amounts so that users of wide verity of capability can utilize them. Each card will be bearing credit equivalent to its' cost.

The customers buy products through the web server of the service provider for making on-line transactions. The web server prompts the customers to enter their card number, password and card provider id. It then sends a request for crediting money to the card provider company. Where it is processed and a notification is made to the service provider about the customer balance and after getting clearance, the service provider completes the rest of the transaction, which includes keeping the record of the transaction in the form of a digital signature [6] and requesting the deduction of credit in card sellers' database. A replica of the digital signature is stored on the credit card provider side so that it may check this against the demand for money by the service provider at the end of the month.

A firewall will be in place between the card provider and the service provider to stand against possible threats of hacking and to prevent the intruders and also secured communication between servers are to be ensured.

## 5. ARCHITECTURE FOR IMPLEMENTATION OF THE FRAMEWORK

The implementation of the proposed framework is simple. The architecture of the total implementation for the proposed framework can be pictorially depicted as Figure 1:

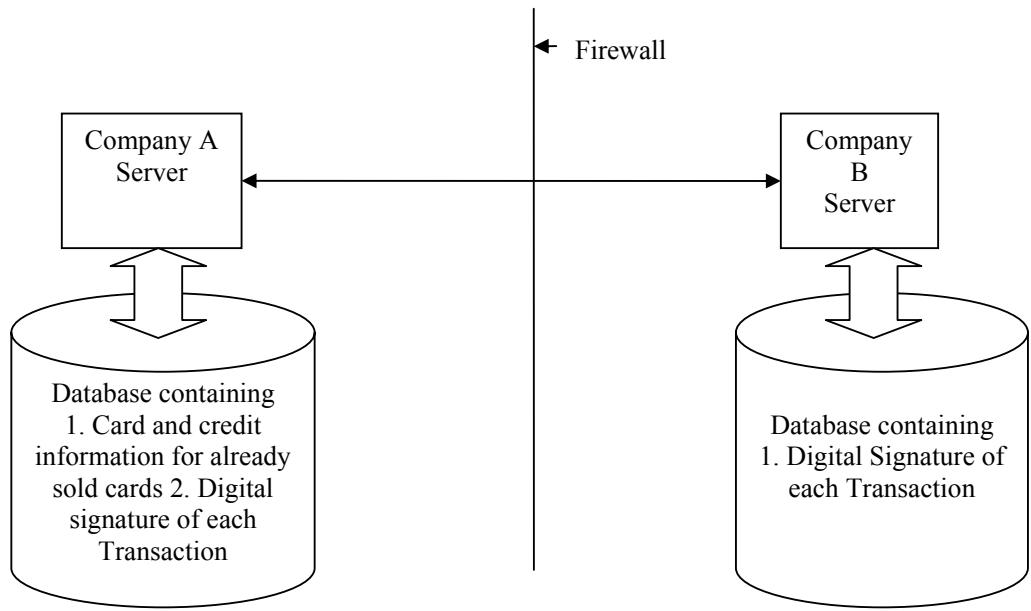

**Figure 1:** The framework for the proposed on-line card based electronic payment system with easy security incorporation.

Here suppose company A prepares a Card according to the defined format and have sold it to a customer. The customer is buying product through a web server which is residing in company B. Whenever a request for crediting money comes the user needs to enter his card's secret number/text, password and the provider's id. Then this information along with the amount requested is passed to the server residing in Company B. It then communicates with the server of company A listening on the predefined port. Server of A then looks into the database and check for the availability of user's credits. If it is sufficient, the server of A notifies B that the requested amount is available. If B agrees then A credits the amount in its server. This transaction takes place along with digital signature. At the end of the month B demands the money from A supplying the digital signature. Since company A is the card provider; it may charge a little percentage as the provider of the card. But that charge will be minimal as because the cost of preparing the card is very low.

## 6. ADVANTAGES OF THE PROPOSED FRAMEWORK

In consideration to the implementation in developing countries, the proposed framework has got a number of added benefits stated as follows:

- This framework can deal with micro transactions.

- As this framework for electronic payment system does not require an on-line third-party connection for each transaction to different banks, there is no threat for processing bottlenecks.
- The customers and the merchants do not require to be registered to a third party. So, there is no registration cost for the buyers and sellers.
- There is no performance bottleneck as with the credit card processing using an on-line third-party, in the verification phase.
- The proposed system is not a candidate for single point of failure as an OTTP.
- The costs of making cards are very low. Hence, the charge for preparing the card will be pretty low too. So, small companies of the developing countries can become interested for spreading their business in internet using this framework. The procedure of transaction is also less tricky to be understood by majority.
- The way of buying and using the cards will be very easy for the users as they are already familiar with the procedure for dialup prepaid internet account recharging and so on.
- The cards may range from $1 to $1000. So, people having different level of capability and need can easily satisfy their necessity.

## 7. IMPLEMENTATION ISSUES

For implementing the framework mainly we are in need of two servers. The one residing in the end of vendor should be capable of establishing secure communication with consumers' browser and communicating with underlying databases holding the digital signature. Another one residing at the end of card producer should be capable of handling secure credit requests and should be equipped with the ability to communicate with the database storing digital signature and sold cards' information. Actually, the questions can definitely arise why we have not generated any implementation result or comparisons. In fact, the proposed framework is thought to be implemented in the developing countries where prepaid card system is operational for the internet browsing. Hence, the applicability and suitability are not the matters of questions, we presume. The design of the server and other issues are just a matter of secure programming and this framework can be used with full commercial benefit. As we are not actually arguing that this framework is better then others in performance or anything, we have not shown any simulation result. Rather we are arguing on the fact that this easy and timely framework can be good alternative for the developing countries to boost up their electronic businesses.

## 8. CONCLUSION

From our point of view the proposed electronic payment system is practically superior to those of the existing ones, when the question for implementing the existing Electronic Payment System in the developing country comes. As it is cheaper and incorporates security within the framework, the proposed framework for on-line prepaid card-based electronic payment system is expected to be suitable for the third world developing countries. So, we conclude with the thought that, the electronic payment system in the developing countries will be profitable and will expectedly open up a new era of different level of e-commerce, if the authorities of the developing countries exploit the proposed framework as the basis for on-line electronic payment in their countries.